\documentclass[sigconf,screen]{acmart}
\usepackage{multirow}

\copyrightyear{2024}
\acmYear{2024}
\setcopyright{acmlicensed}\acmConference[WiPSCE '24]{The 19th WiPSCE Conference on Primary and Secondary Computing Education Research}{September 16--18, 2024}{Munich, Germany}
\acmBooktitle{The 19th WiPSCE Conference on Primary and Secondary Computing Education Research (WiPSCE '24), September 16--18, 2024, Munich, Germany}
\acmDOI{10.1145/3677619.3678117}
\acmISBN{979-8-4007-1005-6/24/09}

\begin{document}

\title[Unpacking Approaches to Learning and Teaching Machine Learning]{Unpacking Approaches to Learning and Teaching Machine Learning in K-12 Education: Transparency, Ethics, and Design Activities}
%% If title is too long for the header, please provide a shorter title (must fit on one line, no breaks!):
%% \title[Shorter Title]{Long Title}

%% The "author" command and its associated commands are used to define
%% the authors and their affiliations.
%% Of note is the shared affiliation of the first two authors, and the
%% "authornote" and "authornotemark" commands
%% used to denote shared contribution to the research.
\author{Luis Morales-Navarro}
\email{luismn@upenn.edu}
\orcid{0000-0002-8777-2374}
\author{Yasmin B. Kafai}
\email{kafai@upenn.edu}
\orcid{0000-0003-4018-0491}
\affiliation{%
  \institution{University of Pennsylvania}
  \city{Philadelphia}
  \state{Pennsylvania}
  \country{USA}
}

%% If the list of authors is too long for the header, please provide a more
%% concise list:
%\renewcommand{\shortauthors}{Trovato and Tobin, et al.}

%% The abstract is a short summary of the work to be presented in the paper.
\begin{abstract}
   In this conceptual paper, we review existing literature on artificial intelligence/machine learning (AI/ML) education to identify three approaches to how learning and teaching ML could be conceptualized. One of them, a data-driven approach, emphasizes providing young people with opportunities to create data sets, train, and test models. A second approach, learning algorithm-driven, prioritizes learning about learning algorithms. In addition, we identify efforts within a third approach that integrates the previous two. In our review, we focus on unpacking how the approaches: (1) glassbox and blackbox different aspects of ML, (2) build on learner interests and provide opportunities for designing applications, (3) integrate ethics and justice. In the discussion, we address the challenges and opportunities of current approaches and suggest future directions for the design of tools and learning activities.  
\end{abstract}

%%
%% The CCSXML block below contains important meta data for the digital
%% library. Please generate adequate meta data for your paper at
%% http://dl.acm.org/ccs.cfm and replace the whole CCSXML block.
%% Also, the tool generates the CCS concepts part of the paper, thus also
%% the ccsdesc commands below the CCSXML block should be replaced accordingly.
%%

\begin{CCSXML}
<ccs2012>
   <concept>
       <concept_id>10003456.10003457.10003527.10003541</concept_id>
       <concept_desc>Social and professional topics~K-12 education</concept_desc>
       <concept_significance>500</concept_significance>
       </concept>
   <concept>
       <concept_id>10003456.10003457.10003527.10003539</concept_id>
       <concept_desc>Social and professional topics~Computing literacy</concept_desc>
       <concept_significance>300</concept_significance>
       </concept>
 </ccs2012>
\end{CCSXML}

\ccsdesc[500]{Social and professional topics~K-12 education}
\ccsdesc[300]{Social and professional topics~Computing literacy}

%%
%% Keywords. The author(s) should pick words that accurately describe
%% the work being presented. Separate the keywords with commas.
\keywords{machine learning, computing education, artificial intelligence, k-12, algorithmic justice, ethics}

%% This command processes the author and affiliation and title
%% information and builds the first part of the formatted document.
\maketitle

\section{Introduction}
While researchers have been investigating how to introduce young people to Artificial Intelligence/Machine Learning (AI/ML) ideas for decades \cite{goldstein1977artificial, kahn1977three}, it is only during the last few years that AI/ML education has gained momentum \cite{kahn_constructionism_2021}. This momentum has been a product of the comeback of machine learning methods, this time powered by increasing computing capacity and increasingly large datasets \cite{touretzky_envisioning_2019}, the establishment of guidelines and principles \cite{touretzky_envisioning_2019}, and the design of tools that enable novices to create models with small data sets \cite{carney_teachable_2020, druga2018growing}. Today, in light of the popularization of large language models and generative models, young people interact with ML every day; governments call for increasing AI/ML education \cite{cardona2023artificial, Eu20}, curriculum providers update their offerings \cite{codeOrg23}, and  teachers scramble to integrate AI/ML content in their classes.  

While principles, guidelines, and considerations serve as guiding frameworks for the design and implementation of learning activities, in practice, these may not always be enacted. Designing ML learning activities requires making decisions about what to glassbox and blackbox, that is deciding on the concepts and practices to prioritize and how to scaffold learners to engage with such a complex topic. Amidst the flurry of activities it is unclear what learning and teaching ML actually looks like, which approaches are taken for different age groups, how learners engage and how critical and ethical issues about ML are being addressed. To address these issues, in this paper, we take a bottom up approach to study the ways in which ML is being introduced to novices. Like Grover \cite{grover2024teaching}, we focus on efforts that foster learning about ML in K-12 and not using ML applications for learning. 

In this conceptual paper, we review the literature on ML education and identify three approaches for learning and teaching ML. A data-driven approach emphasizes facilitating opportunities for youth to produce and curate data sets, as well as to train and evaluate models. A second approach, learning algorithm-driven, places emphasis on fostering learners’ understanding of the inner workings of commonly used learning algorithms. These two are not mutually exclusive, a third approach integrates them. We evaluate these approaches with the following research questions: How are current efforts in ML education in K-12 glassboxing and blackboxing ML content? How are ethics and justice issues integrated into different approaches to ML education? How do different approaches to ML education provide opportunities for designing applications? In the discussion, we address the challenges and opportunities of current approaches and suggest future directions for the design of learning activities. 

\section{Background}

\subsection{ML Education in K-12}
 In their conceptualization of the five big ideas for AI education (see Figure 1), Touretzky et al. \cite{touretzky_envisioning_2019} highlight the role of machine learning methods as the driver of AI adoption over the past fifteen years. Machine learning enables algorithmic systems to learn behaviors without explicit programming \cite{tedre_ct_2021}. This functionality is achieved by a learning algorithm modifying the internal representations of a reasoning model (e.g., decision tree or neural network), allowing it to learn new behaviors. To effectively narrow the algorithm's choices, the reasoning model requires a large amount of training data. Once the machine learning algorithm has created the reasoner, it can be used to solve problems and make decisions on new data \cite{touretzky2023machine}. As such, machine learning provides numerous opportunities for K-12 students to experiment with training data, learning algorithms, and the design of applications.

\begin{figure}
    \centering
    \includegraphics[width=1\linewidth]{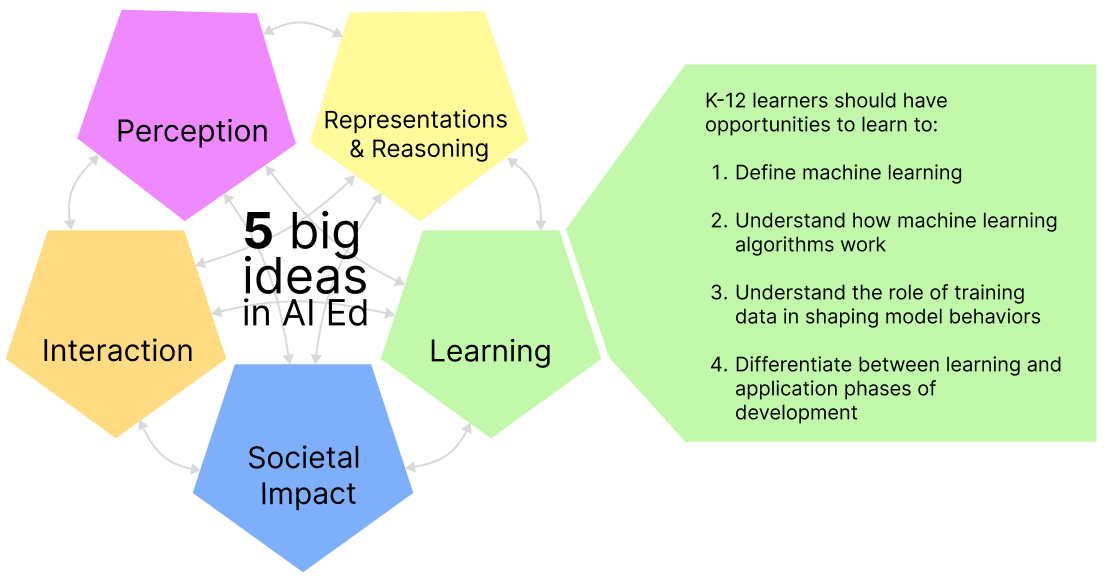}
    \caption{Diagram of the 5 big ideas in AI education proposed by Touretzky et al. \cite{touretzky_envisioning_2019} and (in green box) details about what ML learning activities should promote \cite{touretzky2023machine}.}
    \label{fig:enter-label}
\end{figure}

More recently, Touretzky and colleagues \cite{touretzky2023machine} emphasize that k-12 learners should have opportunities to learn to (1) define machine learning, (2) understand how machine learning algorithms work, (3) understand the role of training data in shaping model behaviors, and (4) differentiate between learning and application phases of development. To do so, they highlight the importance of designing learning activities that support students in building models, experimenting, and creating applications. For example, this may involve having learners train models, learn new concepts from labeled data, construct decision trees with labeled data, simulate how a neural network learns by adjusting its weights, explore historical datasets, and train models based on real-world datasets. Designing learning activities requires making important decisions on what aspects of ML to engage students with, how to scaffold learning about complex concepts and processes, and what tools to use in the classroom. As well as considering how to integrate ethical and critical considerations while learning technical and functional aspects of ML and how to make learning ML relevant to students interests and everyday lives. 

Furthermore, as Tedre and colleagues \cite{tedre_ct_2021} argue, the integration of machine learning into computing education poses challenges to computational thinking practices that have been adopted in the past fifteen years. This is a product of the opacity of ML algorithms, the data driven and data intensive nature of ML. Similarly, Shapiro and Tissenbaum highlight that in contrast to traditional paradigms in computing education, ML is much more empirical, involving learners in conducting experiments, formulating hypotheses, and evaluating models based on predictions \cite{shapiro2019new}. As such, designing learning activities requires making important decisions on how to address the opacity of ML systems, how to engage students in training and evaluating models, and finding accessible entry points.

In reviewing the literature it becomes clear that students should have a conceptual understanding of ML systems in terms of how both data learning algorithms shape model behaviors. Furthermore, students should understand the potential harmful biases and societal implications of ML systems, as well as how to design and deploy them ethically. Finally, students should be able to apply ML to real-world problems in order to prepare for future careers that may involve working with ML technologies and leveraging them to address real-world challenges.

\subsection{Blackboxing and Glassboxing}
In the learning sciences, the question of what to make visible in learning computing has been widely discussed in relation to blackboxing and glassboxing \cite{hmelo1996black, resnick1999beyond}. These concerns have been discussed since the early days of symbolic AI education \cite{goldstein1977artificial, peelle1974computer}. Goldstein and Papert \cite{goldstein1977artificial}, for instance, argued for a glassbox approach to enable learners to “look inside the box, to ask how it works.” In the case of contemporary ML education, this is particularly relevant because of the opacity and blackboxed nature of models and data discussed above \cite{broll_beyond_2022, hitron_can_2019}.

Hmelo and Guzdial \cite{hmelo1996black} highlighted that when dealing with complexity in learning-by-doing tasks, blackbox and glassbox scaffoldings support students in different ways. Blackbox scaffoldings support learners to perform and complete tasks; without them, learners would otherwise be unable to complete a task. For instance, in ML education blackboxing can be productive to enable k-12 learners to train and test models without requiring them to have advanced math knowledge and programming experience. Glassbox scaffoldings, on the other hand, provide support that “allow the learner to look inside,” making processes explicit. In this sense, the scaffoldings are not permanent but are meant to fade. They discuss that the choice between blackbox and glassbox scaffoldings is a design decision that depends on the goals of learning activities. Furthermore, Resnick, Berg, and Eisenberg \cite{resnick1999beyond}, discuss the importance of moving beyond blackboxes to bring back transparency to how computing technologies operate. Their discussion centers on how most computing tools used in education are blackboxed with “their inner workings often hidden and thus poorly understood by their users” and the importance of designing tools that make visible their inner workings and support learners to create personally relevant projects.

\subsubsection{Designing Applications}

As Resnick and colleagues \cite{resnick1999beyond} argue, tools that glassbox the inner workings of computing can support learners in designing applications that would otherwise be inaccessible. Designing applications in computing education has a long history that goes back to studies on LOGO \cite{harel1991children, kafai2012minds}. This approach of having K-12 students create projects that relate to their interests, instead of completing pre-determined problem sets, has gained traction over the past decade \cite{waite2021teaching, oleson2020role}.  In ML education, for instance, learning activities may blackbox and glassbox different aspects of the ML pipeline to support students to move beyond just training models in isolation to create models to be used in fully developed applications.

\subsubsection{Algorithmic Justice and Ethics}

More recently, Dixon, Hsi and Oh \cite{dixon2022unblackboxing} revisited the ideas of providing blackbox and glassbox scaffoldings to propose that transparency should not be solely focused on technical disciplinary knowledge and practices; instead, it should also consider the histories, externalities, and possible futures of computing technologies. Here it is important for youth to engage with ideas of algorithmic justice. Algorithmic justice considers that ML systems have implications on  individuals and communities that could perpetuate harm in unjust ways, disproportionately impacting vulnerable populations \cite{birhane2021algorithmic}. As such, making ML blackboxes transparent for learners involves providing scaffolds to think through how functional and critical issues are intertwined, consider the possible ways in which ML applications could be used, and take into account their ethical, social, and environmental implications.

\section{Methods}

In this section we provide the reader with details about how we selected and analyzed existing studies to conceptualize the data-driven and learning algorithm-driven approaches. Since the goal of this paper is to conceptualize different ways in which ML education is enacted through learning and teaching we decided to conduct a narrative critical review of the literature \cite{hempel2020conducting}. Our analysis was guided by the following research questions: How are current efforts in ML education in K-12 glassboxing and blackboxing ML content? How are ethics and justice issues integrated into different approaches to ML education? How do different approaches to ML education build on learner interests and provide opportunities for designing applications?  

We began our literature review by searching for papers on ML education in the K-12 context in proceedings and journals associated with or sponsored by ACM’s Special Interest Group in Computer Science Education \footnote{Australasian Computing Education, ACM Conference on International Computing Education Research, ACM Conference on Innovation and Technology in Computer Science Education, Koli Calling International Conference on Computing Education Research, ACM Technical Symposium on Computer Science Education, Workshop in Primary and Secondary Computing Education, ACM Transactions of Computing Education}, ACM’s Special Interest Group in Computer-Human Interaction \footnote{Cognition \& Creativity, CHI Conference on Human Factors in Computing Systems, Computer Supported Collaborative Work, Designing Interactive Systems, Interaction Design and Children, Conference on tangible, embedded, and embodied interaction, ACM symposium on User interface software and technology}, AAAI \footnote{AAAI Symposium on Educational Advances in Artificial Intelligence}, the International Society of the Learning Sciences \footnote{Information and Learning Sciences, Journal of Learning Sciences, International Conference of the Learning Sciences, ISLS Rapid Reports}, Computers \& Education journals\footnote{Computer \& Education, Computers \& Education: Artificial Intelligence, Computers \& Education: Open}, IEEE education \footnote{IEEE Access, IEEE International Conference of Educational Innovation Through Technology, IEEE Frontiers in Education Conference, IEEE International Conference on Advanced Learning Technologies,  IEEE Integrated STEM Education Conference}, and other education, human-computer interaction, and AI/ML journals and conferences. \footnote{British Journal of Education Technology, AI Matters, International Journal of Child-Computer Interaction, Interactive Learning Environments, IEEE Symposium on Visual Languages and Human-Centric Computing,  KI-Künstliche Intelligenz} We also included papers mentioned in other reviews of the literature \cite{long_what_2020, druga_landscape_2022, marques2020teaching, rauber2022assessing, sanusi2023systematic, rizvi2023artificial}. Initially, we identified 206 studies that were relevant to our study by reading their titles and abstracts. Following, we read through the full text of these 206 papers to determine their inclusion and identified 72 different studies that explicitly described ML learning activities or curricula for K-12 students to learn about ML models. We excluded studies that focused on supporting students to learn how to use ML applications (e.g., studies on the use of large language models in programming assignments), studies that focused solely on the discussion of ethical issues related to ML without providing opportunities for learners to engage with functional aspects of ML, and studies on teachers that did not explicitly describe the curriculum teachers implemented with students. 

We then analyzed the papers in five steps. First we read the papers and kept track of how the interventions and learning activities described in the studies introduced ML, what the learning activities looked like and what and how concepts and ideas (including both technical and critical/ethics) were included. Second, we reviewed our notes and through iterative discussion grouped studies together depending on whether they prioritized data or learning algorithms. Third, we systematically coded all 72 studies using binary codes to apply the same criteria to all papers \cite{hempel2020conducting} by indicating if they approached ML from a data-driven perspective, a learning algorithm perspective, if the activities described in the studies involved students in creating applications, personally relevant projects, discussions about justice and ethics and evaluating ML models. We also noted if the studies introduced new tools and curricula. Fourth, we further analyzed papers that approached ML from both data and learning algorithm perspectives and grouped them into three categories: mix and match, data driven with algorithm sprinkles, and algorithmic-driven with data sprinkles. Finally we wrote memos describing what studies in each approach black and glassboxed, the curricula and tools presented, and how these involved students in making applications, addressing issues of justice and ethics. The literature review was carried out by the first author, who sought the second author's input and expertise during weekly meetings.  

\section{Findings}

Overall we identified 72 papers that deal with learning and teaching about ML in K-12 education. Based on our analysis we identified three distinct approaches to ML education in K-12 that, through the design of learning activities, blackbox and glassbox different aspects of ML (see Table 1). The most common approach, the data-driven approach, focuses on having students create datasets to train and test models (see Figure 2) but blackboxes the learning algorithms. The second approach engages students with learning algorithms and blackboxes the datawork involved in the ML pipeline. A third integrative approach involves a combination of data-driven and learning algorithm driven learning activities.

% Please add the following required packages to your document preamble:
% \usepackage{multirow}
\begin{table*}[]
\caption{Approaches to learning and teaching ML}
\scalebox{0.85}{
\begin{tabular}{llll}
\hline
\multicolumn{2}{l}{\textbf{Approaches}} &
  \textbf{Studies} &
  \textbf{Age groups}* \\ \hline
\multicolumn{2}{l}{\begin{tabular}[c]{@{}l@{}}\textbf{Data-driven:} glassboxes how data shapes \\ model performance and blackboxes the \\ role of learning algorithms in the ML \\ pipeline.\end{tabular}} &
  \begin{tabular}[c]{@{}l@{}}Agassi et al. \cite{agassi_scratch_2019}, Arastoopour Irgens et al. \cite{arastoopour_irgens_characterizing_2022}, \\ Bilstrup et al. \cite{bilstrup_staging_2020}, Castro et al. \cite{castro_ai_2022}, Dietz et al. \cite{dietz_artonomous_2022},\\ Druga \& Ko \cite{druga_how_2021}, Druga et al. \cite{druga_family_2022}, Dwivedi et al., \cite{dwivedi_exploring_2021},\\ Henry et al. \cite{henry_teaching_2021}, Hitron et al. \cite{hitron_can_2019, hitron_introducing_2018}, Hjorth \cite{hjorth_naturallanguageprocesing4all_2021},\\ Jiang et al. \cite{jiang_empirical_2022}, Jordan et al.\cite{jordan_poseblocks_2021}, Kaspersen et al. \cite{kaspersen_machine_2021},\\ Katuka et al. \cite{katuka2023summer, katuka2024integrating}, Krakowski et al. \cite{krakowski_authentic_2022}, Lee et al. \cite{lee_ai-infused_2021},\\ Lin et al. \cite{lin_zhorai_2020}, Martin et al. \cite{martin2024chemaistry}\, Morales-Navarro et al.\cite{morales2024not},\\ Ng et al.\cite{ng_using_2022}, Park et al. \cite{park_designing_2021}, Rodríguez-García et al. \cite{rodriguez-garcia_evaluation_2021},\\ Sabuncuoglu \cite{sabuncuoglu_designing_2020}, Song et al. \cite{song_paving_2022}, Toivonen et al. \cite{toivonen_co-designing_2020, toivonen_interacting_2022}, \\ Tseng et al. \cite{tseng_plushpal_2021, tseng2024co, tseng2023collaborative} , Vartiainen et al. \cite{vartiainen_learning_2020, vartiainen_machine_2020, vartiainen_machine_2021}, Williams et al. \cite{williams_is_2019, williams_popbots_2019}, \\ Zhu \& Van Brummelen \cite{zhu_teaching_2021}, Zimmermann-Niefield et al. \cite{zimmermann-niefield_youth_2019,zimmermann-niefield_youth_2020}\end{tabular} &
  \begin{tabular}[c]{@{}l@{}}Early childhood - K: \\ 2 studies \\ Primary school: \\ 21 studies\\ Middle school/early secondary: \\ 17 studies\\ High school/late secondary: \\ 10 studies\end{tabular} \\ \hline
\multicolumn{2}{l}{\begin{tabular}[c]{@{}l@{}}\textbf{Learning algorithm-driven:} glassboxes \\ how learning  algorithms work and \\ blackboxes  the datawork involved in \\ the ML pipeline\end{tabular}} &
  \begin{tabular}[c]{@{}l@{}}Akram et al. \cite{akram_towards_2022}, Alvarez et al. \cite{alvarez_socially_2022}, Ma et al. \cite{ma2023developing}, Norouzi et al. \cite{norouzi_lessons_2020}, \\ Priya et al.\cite{priya_ml-quest_2022}, Quiroz et al. \cite{quiroz2024scratch}, Sperling \& Lickerman \cite{sperling_integrating_2012}\end{tabular} &
  \begin{tabular}[c]{@{}l@{}}Middle school/early secondary: \\ 2 studies\\ High school/late secondary:\\ 5 studies\end{tabular} \\ \hline
\multirow{3}{*}{\textbf{Integrative}} &
  \begin{tabular}[c]{@{}l@{}}\textbf{Mix and Match:} integrates \\ both approaches by having \\ different learning activities \\ that focus on either approach\end{tabular} &
  \begin{tabular}[c]{@{}l@{}}Broll et al. \cite{broll_beyond_2022}, Fernández-Martínez et al. \cite{fernandez-martinez_early_2021}, Lee et al.,  \cite{lee_preparing_2022, lee_developing_2021},\\ Long et al. \cite{long_co-designing_2021, long_designing_2019, long_role_2021, long_family_2022, long2023fostering}, Mahipal et al. \cite{mahipal_doodleit_2023},\\ Mariescu-Istodor \& Jormanainen \cite{mariescu-istodor_machine_2019}, Reddy et al. \cite{reddy_levelup_2022},\\ Van Brummelen et al., \cite{van_brummelen_alexa_2021, van_brummelen_teaching_2021},  Voulgari et al.\cite{voulgari_learn_2021}\end{tabular} &
  \begin{tabular}[c]{@{}l@{}}Primary school:\\ 4 studies\\ Middle school/early secondary:\\ 11 studies \\ High school/late secondary:\\ 7 studies\end{tabular} \\ \cline{2-4} 
 &
  \begin{tabular}[c]{@{}l@{}}\textbf{Data-driven with learning} \\ \textbf{algorithm  sprinkles:} \\ involves data-driven activities\\ while also explaining the \\ learning algorithms used in \\ models through lectures,\\ discussions and videos\end{tabular} &
  \begin{tabular}[c]{@{}l@{}}Ali et al., \cite{ali_what_2021},  Buxton et al. \cite{buxton2024foundations}, Guerreiro-Santalla et al. \cite{guerreiro-santalla_smartphone-based_2022}\\ Kaspersen et al. \cite{kaspersen_high_2022}, Ng et al. \cite{ng_motivating_2021}, Shamir \& Levin \cite{shamir_teaching_2022}\\ Williams et al., \cite{williams_teacher_2021}\end{tabular} &
  \begin{tabular}[c]{@{}l@{}}Primary school: \\ 1 study\\ Middle school/early secondary: \\ 4 studies \\ High school/late secondary: \\ 4 studies\end{tabular} \\ \cline{2-4} 
 &
  \begin{tabular}[c]{@{}l@{}}\textbf{Learning algorithm-driven} \\ \textbf{with data sprinkles:} focuses \\ on learning algorithms and \\ includes discussions about how \\ data influences model behavior\end{tabular} &
  \begin{tabular}[c]{@{}l@{}}Reddy et al. \cite{reddy_text_2021}, Wan et al.\cite{wan_smileycluster_2020}, Zhou et al. \cite{zhou_scaffolding_2021}\end{tabular} &
  \begin{tabular}[c]{@{}l@{}}Middle school/early secondary: \\ 1 study\\ High school/late secondary: \\ 2 studies\end{tabular} \\ \hline
\multicolumn{4}{r}{*Some studies involved students in more than one age group.}
\end{tabular}
}
\end{table*}

\begin{figure}
    \centering
    \includegraphics[width=\linewidth]{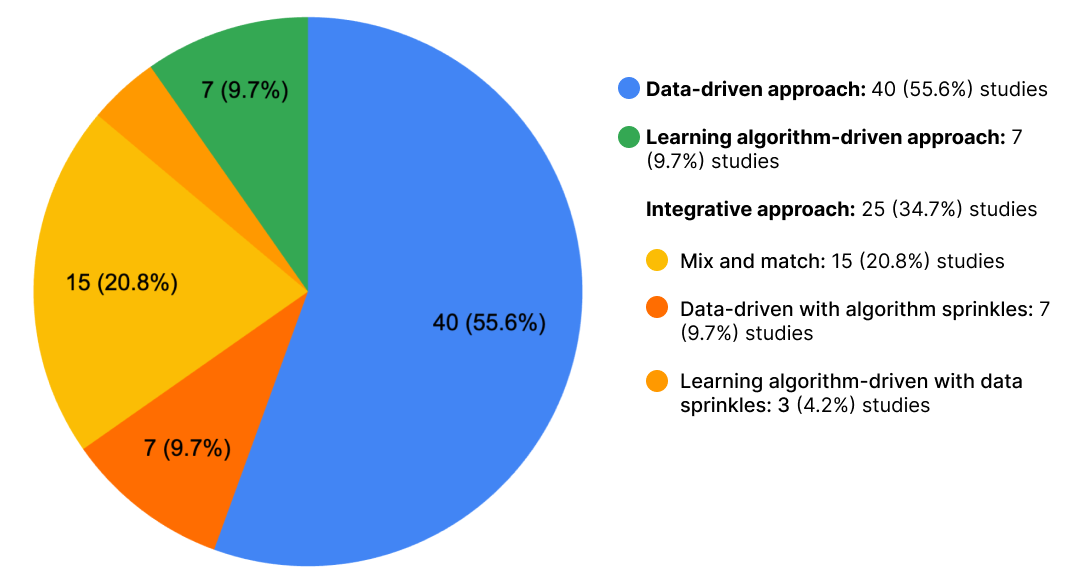}
    \caption{Distribution of studies by approach}
    \label{fig:enter-label}
\end{figure}

\subsection{Data-driven approach}

The most common approach to AI/ML education in K-12 is data-driven, having students build datasets to train models. This approach glassboxes how data shapes model performance and blackboxes the role of learning algorithms in the ML pipeline. We reviewed 40 different studies that took a data-centric approach. Most studies within the past five years have focused on having young people create and label datasets for classification tasks (e.g., \cite{ng_using_2022, jordan_poseblocks_2021, druga_family_2022, arastoopour_irgens_characterizing_2022, hjorth_naturallanguageprocesing4all_2021}. This approach engages young people with what Fiebrink \cite{fiebrink2019machine} calls "Interactive Machine Learning" which involves iteratively ideating a system and its desired behavior, implementing it, observing its behavior, and comparing the observed behavior with the desired behaviors. Here users create small data sets that can be easily and quickly refined and edited to retrain models and improve their performance \cite{zimmermann-niefield_youth_2019}.  However, Hitron et al. \cite{hitron_can_2019} show that to build basic conceptual understanding of machine learning through a data-driven approach, solely labeling data or just evaluating models is not enough. In a study with children they found that engaging with the full pipeline of creating a dataset, training and testing a model can lead to significant improvements in conceptual understanding when compared with just labeling data or evaluating outcomes. More recently, researchers have articulated data design practices for novices building ML models based on expert practices \cite{tseng2024co}. These include 1) incorporating dataset diversity, 2) evaluating model performance and its relationship to data, 3) balancing datasets, and 4) inspecting for data quality. The data-driven approach promotes what Vartianen and colleagues \cite{vartiainen_machine_2021} call “data-driven reasoning and design” which involves thinking about decisions made in the design of datasets to explain the behavior of machine learning systems. As such in this approach model behaviors are ascribed to the design of datasets. 

This approach has been used in diverse contexts: in and out of school, with very young children \cite{vartiainen_learning_2020, williams_popbots_2019} and adolescents (e.g., \cite{tseng2024co, sabuncuoglu_designing_2020}), in computing classes and workshops \cite{arastoopour_irgens_characterizing_2022} and in other disciplines such as social studies \cite{hjorth_naturallanguageprocesing4all_2021} or chemistry \cite{martin2024chemaistry}. For example Lin and colleagues \cite{lin_zhorai_2020} involved young children in designing a conversational agent by providing data related to specific animals. Sabuncouglu and colleagues \cite{sabuncuoglu_designing_2020}, on the other hand, designed and implemented a year-long curriculum to introduce adolescents to machine learning, in which they designed data sets for rock, paper, scissors classifiers, and a personal project to address environmental issues. One study went beyond classification tasks by having children provide data (melodies) for a generative music model \cite{williams_popbots_2019}. Other studies built on unplugged computing activities, using a card game for young people to reflect on the context of data, data inputs and data ethics \cite{bilstrup_staging_2020}. 

In the following subsections we describe how studies in the data-driven approach engage learners in making models in the context of designing applications, propose different tools and curricula, involve learners in testing, and address algorithmic justice issues. 

\subsubsection{Tools \& Curricula}
Studies that approached ML education from a data-driven perspective introduced tools to support learners in building models and presented curricular interventions.

Twenty six studies introduced new tools that enable learners to build and create models. These included extensions of Scratch to support face, body, and hand recognition and the possibility of including Teachable Machine models \cite{jordan_poseblocks_2021, rodriguez-garcia_evaluation_2021}, novel tools for using natural language processing in social studies classrooms \cite{hjorth_naturallanguageprocesing4all_2021, jiang_empirical_2022}, and gesture classification tools. One tool involved a physical tangible artifact that could be used by children to create datasets using drawings, train models and test them \cite{kaspersen_machine_2021}. Hjorth \cite{hjorth_naturallanguageprocesing4all_2021} presented Text Tagging, an application for learners to use NLP techniques on social media posts in their social studies classes. Gesture classification tools included Gest \cite{hitron_introducing_2018}, PlushPal \cite{tseng_plushpal_2021} and AlpacaML \cite{zimmermann-niefield_youth_2019} which all supported learners in building and testing ML models that use accelerometer sensor data. Despite their similarities, each of these tools supports different kinds of engagement with gesture classification from supporting young children in telling stories about their plushies \cite{tseng_plushpal_2021} to teenagers in creating sports related projects \cite{zimmermann-niefield_youth_2019}. More advanced tools, such as StoryQ \cite{jiang_empirical_2022}, also support learners to visualize and explore patterns in data. Unplugged tools included a card game about data for ML models, with cards that determined the context of data, data inputs, and prompts about how data is collected and its ethics \cite{bilstrup_staging_2020}. 

Nineteen studies presented curricular activities to engage students with data-driven ML topics in distinct ways. Ng et al. \cite{ng_using_2022} for example presented a curriculum on storytelling and ML and Hjorth \cite{hjorth_naturallanguageprocesing4all_2021} introduced ML activities for social studies. Other curricula took a problem-based approach to have youth address socio-technical problems using ML \cite{krakowski_authentic_2022}. Some efforts included curricular activities for at-home learning with families \cite{druga_family_2022, tseng2023collaborative}. Scaffolds for learners included activity cards, sample projects and worksheets \cite{jordan_poseblocks_2021}. While some efforts were short after school activities \cite{arastoopour_irgens_characterizing_2022} others involved year long curricula \cite{rodriguez-garcia_evaluation_2021}.

\subsubsection{Making models in the context of designing applications}

Of the 40 studies we reviewed, 19 studies had students train models in the context of building applications and 24 centered on building models in isolation. Building applications in this context involved creating datasets and training models and using those models in applications usually designed by learners themselves. As such, learners had to engage both with machine learning as well as traditional computing education practices and concepts (see \cite{tedre_ct_2021} for a discussion of the difference between traditional computing and ML education). 

Sixteen of these studies involved students in model building in the context of creating personally relevant projects. Poseblocks \cite{jordan_poseblocks_2021} for example, involved students in creating AR filters related to their personal interests. Several studies had students build models to use in personally relevant Scratch projects \cite{zimmermann-niefield_youth_2020, ng_using_2022, jordan_poseblocks_2021}. Examples here included children making a project to help their siblings learn their ABCs \cite{jordan_poseblocks_2021}. However, as Zimmermann-Niefield and colleagues \cite{zimmermann-niefield_youth_2020} noted sometimes the projects and models can be incoherent, in the case of their study, with classified gestures having little relationship to narratives and characters of Scratch projects. 

Other studies involved learners in building data sets, training and testing models and brainstorming applications in which these could be used \cite{druga_family_2022, toivonen_co-designing_2020, vartiainen_learning_2020, vartiainen_machine_2021}. Here learners created paper prototypes of applications that could use their models. In a couple of these studies the prototypes were then developed by software engineers to have learners test them \cite{vartiainen_machine_2020, vartiainen_machine_2021}. While this is not scalable and applicable in most k-12 classrooms, it provided a rich opportunity for students to see how the models they trained could be used in real applications.

\subsubsection{Testing}
Most studies focused on having learners prepare training datasets for models with little attention paid to creating testing datasets and testing. Indeed, most testing efforts presented in current research involved live-testing, that is using a trained model to classify new inputs in real time \cite{toivonen_co-designing_2020, vartiainen_machine_2021, druga_how_2021}. Studies show that live-testing can be beneficial for youth to take perspective on their models, build hypotheses, and iteratively improve model performance \cite{morales2024not, vartiainen_machine_2021, druga_how_2021}; yet testing activities seem to work better when explicitly scaffolded by instructors \cite{zimmermann-niefield_youth_2020}. Engaging with testing data sets was only present in a few studies. In Popbots \cite{williams_popbots_2019}, for example, young children experimented with different training datasets to see their impact on accuracy in classifying a testing set. In studies by Tseng et al. \cite{tseng2023collaborative, tseng2024co} youth created testing data sets and determined the accuracy of their models by class or label. Other studies mention evaluation more broadly. Krakowski et al. \cite{krakowski_authentic_2022}, for instance, framed evaluation as a process that involves: determining if ML is appropriate for the task at hand, reviewing and questioning the data used to train a model, considering the affordances and limitations of the design of ML systems, and taking into account the foreseeable impact of ML decisions. 

\subsubsection{Algorithmic justice and ethics}
Few studies that took on a data-driven approach (only 12 out of 40) addressed issues of algorithmic justice.  At the same time, while the studies acknowledged the importance of learning about ethics and reducing algorithmic harms, these topics were often presented as something that should be learned separate from skills and concepts. Here, ethics were discussed in relation to commercial applications instead of being applied to learner-designed applications. For example, some studies used ethical matrices for children to redesign YouTube \cite{sabuncuoglu_designing_2020} or fostered discussions of ethics while redesigning voice assistants \cite{druga_family_2022}. A common activity in several studies involved watching videos of Joy Buolamwini \cite{jordan_poseblocks_2021, arastoopour_irgens_characterizing_2022}).

Some approaches to issues of justice may be problematic. One study, for example, showed that some children’s concerns may align with long termist sci-fi-inspired ideas \cite{ng_using_2022}. This shows that it is important to scaffold conversations about justice and ethics to ensure learners can have informed discussions. Another issue that may be problematic is the idea of unbiased ML systems, which a teacher in one study voiced \cite{ng_using_2022}. Some studies foster AI4SocialGood ideas (e.g., \cite{krakowski_authentic_2022, arastoopour_irgens_characterizing_2022}), here it is important to avoid falling into the trap of technosolutionism and to ask if ML is necessary to address the problems at hand. 

Some promising studies integrate discussion of justice into the design of applications. Here, Jordan et al. \cite{jordan_poseblocks_2021} had children create ethical matrices for their teachable machine classifiers, and Bilstrup et al. \cite{bilstrup_staging_2020} prompted youth to discuss the ethics of their projects with a card game that poses questions such as: “Are users aware that data is collected about them?" and “Is the use of ML visible in your system?” Lin and colleagues \cite{lin_zhorai_2020} had young children discuss issues of misinformation in relation to a learner-trained conversational agent.

\subsection{Learning algorithm-driven approach}
A different approach centers on glassboxing how learning algorithms work and blackboxing the datawork involved in the ML pipeline. We only identified seven different studies that took a learning algorithm-driven approach. In these studies, learners were introduced to neural networks, nearest neighbor, k closest neighbors, NLP methods, and search algorithms, among others, using off-the-shelf datasets (e.g., datasets from Kaggle). These studies were all conducted with secondary school students in \cite{sperling_integrating_2012} and out of school \cite{norouzi_lessons_2020}. 

While some studies involved students learning about ML in computing settings such as CS classes and summer camps \cite{norouzi_lessons_2020, sperling_integrating_2012, alvarez_socially_2022}, others integrated ML into science curriculum \cite{akram_towards_2022}. For students in computing contexts ML learning activities required some pre-existing knowledge of programming. As such, camps included python bootcamps \cite{norouzi_lessons_2020} or  introductory python activities \cite{alvarez_socially_2022} and classroom interventions involved students with more than 2 years of traditional computing education and knowledge of python \cite{sperling_integrating_2012}. On the other hand, introductory activities in non-computing contexts used block programming environments or no programming at all. Akram et al. \cite{akram_towards_2022}, for instance, presented curricular modules that integrate ML using a block programming environment into early secondary school (grades 6-8) science. Each module involved a science topic, a core ML algorithm, and an algorithmic justice topic. For example one module involved using the breadth-first search algorithm, uniform cost search and  adversarial search for contact tracing in the context of Covid-19 while discussing issues of privacy. Other algorithms included knowledge-based systems, clustering with K-means, decision trees, NLP methods (feature extraction, information retrieval, feature selection, classification). Priya and colleagues \cite{priya_ml-quest_2022}, created a game that provided a conceptual overview of supervised learning, gradient descent, and K-Nearest Neighbor classification without opportunities to build and experiment with models.

\subsubsection{Tools \& Curricula}

Studies involved curricular activities of different length from 90 minutes \cite{akram_towards_2022} to 450 hours \cite{sperling_integrating_2012}. Akram and colleagues' \cite{akram_towards_2022} ML modules for science classrooms were short by design so that they can be integrated when learning about epidemics, natural disasters, and gravity. Norouzi et al. \cite{norouzi_lessons_2020} presented a four week curriculum that included one week of coding exercises with python, one week on NLP techniques, and one week for working on projects, guest lectures and field trips. Sperling’s \cite{sperling_integrating_2012} 450 hour course introduced theoretical ideas of algorithms, machine learning, ML algorithms and provided opportunities for students to create full-fledged final projects. Two of the studies we reviewed introduced novel extensions to the Snap programming environment that enabled students to use and explore ML algorithms \cite{akram_towards_2022, alvarez_socially_2022}.

\subsubsection{Making models in the context of designing applications}

Studies in the algorithm-driven approach involved students in using different ML algorithms and techniques to analyze data and sometimes create applications. Because most studies used off-the-shelf datasets, many of the projects and applications that students worked on were predetermined by curricular designers, instructors, and researchers \cite{akram_towards_2022, norouzi_lessons_2020}. These involved, for example, having students detect breast cancer or malaria \cite{norouzi_lessons_2020}, or build covid contact tracing and disaster risk prevention applications \cite{akram_towards_2022}. Alvarez et al., on the contrary, provided students with open-ended activities in which they could use NLP to analyze any data of their choice from Genius Lyrics, Twitter, and the New York Times APIs. In Sperling \cite{sperling_integrating_2012}, as part of a 450 hours curriculum students created their own application of their choice using an ML algorithm, built a user interface, and examined the performance of their project.

\subsubsection{Testing}

Only one of the studies reviewed reported involving students in testing models. In  Norouzi et al. \cite{norouzi_lessons_2020} on the last day of a workshop, students tested their projects with their classmates with an emphasis on accuracy and identifying ways to increase the accuracy of the models. 

\subsubsection{Algorithmic justice and ethics}
Three studies addressed issues of algorithmic justice through discussions and conversation and not in relation to the projects that students created. Alvarez and colleagues \cite{alvarez_socially_2022} had students discuss how ML is used in the criminal justice system. In Akram et al. \cite{akram_towards_2022} at the end of each module, students had an ethics in AI discussion; the topics covered included privacy, bias, fairness, and automation. Norouzi et al. \cite{norouzi_lessons_2020} included a guest lecture on AI ethics in which they discussed how ML systems can replicate biases present in training data.

\subsection{Integrative approach}
Finally, we identified three different ways in which data and algorithmic driven approaches were integrated. Some studies involved students in learning about how data and learning algorithms together shape model behaviors. Other studies focused on having students complete data-driven activities (such as creating data sets to train and then test models) and provided some explanation or discussion of the learning algorithms used (in the form of lectures and videos). A third group of studies centered on learning algorithms while including discussions and activities about how data influences model behaviors.

\subsubsection{Mix \& match }

We identified 15 different studies that integrated both approaches often having different hands-on learning activities that focus on either approach. Long and Magerko \cite{long_designing_2019, long_co-designing_2021}, for example, involved youth in learning about ML with unplugged activities in museum settings. They integrated both approaches by creating distinct exhibit experiences focused on ML competencies. For example, some exhibits centered on explaining how neural networks work and their representations while others prioritized engaging youth with data and how data shapes system behaviors.

In formal school settings, Lee and colleagues \cite{lee_preparing_2022} presented an ML methods in data science curriculum in which high school students first engaged with a unit focused on data, exploring datasets, considering ethical issues of data production, and analyzing data, building datasets for image classification and later participated in activities designed to learn about perceptrons, neural networks, back propagation, transfer learning, and K nearest networks. Of note data design issues were also integrated in the activities related to learning algorithms as the curriculum focused on the whole ML pipeline. In a different curriculum Lee and colleagues \cite{lee_ai-infused_2021} offered a comprehensive 30 hour introduction to ML for middle schoolers that included both data-driven units and algorithm-driven units. From creating data sets to train classifiers to exploring the structure of neural networks to an introduction to generative adversarial networks. Here every unit also included an ethics component. 

Researchers have also designed tools to support sensemaking of how data and learning algorithms work together in classification tasks. Mahipal et al. \cite{mahipal_doodleit_2023} presented a curriculum for middle schoolers that integrated both data and algorithmic driven approaches. In this curriculum students used DoodleIt, to explore and visualize how data is processed by a neural network. This tool, while classifying drawings in real time, visualizes the application of the kernels of a convolutional neural network to the data by creating feature maps.

\subsubsection{Data-driven with algorithm sprinkles}

Eight studies involved students in data-driven activities (such as creating data sets to train and then test models) while also explaining the learning algorithms used in such models through lectures and videos. In How To Train Your Robot  \cite{williams_teacher_2021}, for example, hands-on learning activities involved creating classifiers for images and text accompanied by explanations of the k-nearest neighbor algorithm and how it is used. Shamir and Levin \cite{shamir_teaching_2022} had students train and test classifiers and later design a logic gate to simulate a (single) artificial neuron. Kaspersen et al. \cite{kaspersen_high_2022}  introduced VotestratesML, a tool that supports students in processing and creating datasets to train models in the context of social studies classes. This data-driven tool enabled learners to choose if they wanted to train their models using k-nearest neighbors or feedforward neural networks, as well as the parameters used in each algorithm (the value of k, the number of training iterations). As such, students could compare the performance of models trained with both algorithms.

\subsubsection{Algorithm-driven with data sprinkles}
Three studies focused on learning algorithms and included discussions and activities about how data influences model behavior. Here for instance, Wan et al. \cite{wan_smileycluster_2020} and Zhou et al.\cite{zhou_scaffolding_2021} had students interact with SmileyCluster, a tool that supports students to make sense of k-means clustering. Here, while students interacted with data, the goal of the activities was to support learners in understanding the k-means clustering method. In another study, Reddy et al. \cite{reddy_text_2021} engaged students with a curriculum that emphasized word embeddings, the k-nearest neighbors algorithm and classification bias. Here students built datasets for the programming activity of their final project, but datasets were not the focus of the intervention.

\section{Discussion}

Through our review of the literature we identified and conceptualized three approaches to learning and teaching ML in K-12. Here we discuss the implications of what we observed in each approach. 
First, we observed that the majority of efforts in ML education have focused on data driven approaches with much less attention given to the role that learning algorithms play in ML model performance. The data driven approach glassboxes the role of training data in shaping model behaviors, current efforts within this approach include learning activities for K-12 students of all ages. On the other hand, all studies in the learning algorithm-driven approach, which glassboxes how learning algorithms work, centered on secondary school students with previous experience with programming and more advanced knowledge of mathematics. This disparity highlights the need to develop learning activities and curricula that make learning algorithms accessible to younger students. 

Second, as Touretzky and colleagues \cite{touretzky2023machine} argue it is important for students to have opportunities to understand ML systems from both data and learning algorithm perspectives, from our analysis we see great potential for further developing the integrative approach beyond mix and matching activities and adding sprinkles here and there. As we saw in our analysis, when learning activities include sprinkles of data or learning algorithm issues, engagement is often limited to discussion and lectures. Mixing and matching by alternating short activities that address how both data and learning algorithms shape model behavior is a good starting point, but we must strive to create activities where students can design datasets and explore how the learning algorithms used to train models shape performance. Here we see the need for better tools that can support both creating training and testing datasets, modifying model parameters and visualizing how learning algorithms work. At the same time, we recognize integrative approaches may be better suited for longer interventions (such as \cite{lee_preparing_2022}) throughout the school year and not one off workshops.

Third, we know that, in computing education, creating applications, particularly personally relevant applications, can provide a context that motivates students to learn computing \cite{waite2021teaching, oleson2020role}. Yet many efforts in ML education prioritize learning about and creating models in isolation without incorporating them into applications. Almost half of the studies in the data driven approach involved having students build applications, providing opportunities for learning both ML and traditional computing education practices and concepts. In the algorithm-driven approach, while some efforts supported students to use learning algorithms with data sets of their own interest, others involved using datasets selected by researchers and instructors, limiting the opportunities to create personally relevant applications.

Fourth, we were not surprised to see that the majority of ML education efforts we reviewed did not include any algorithmic justice or ethics content, as this is also the case in undergraduate education \cite{saltz2019integrating}. Oftentimes, when ethics or issues of justice were mentioned in learning activities these were disconnected from the technical and functional aspects of ML, for example simply showing the trailer of a documentary or having a discussion about the ethics of self-driving cars in a workshop where nothing else had to do with self-driving cars. We argue that issues of justice and ethics should be addressed in conjunction with technical issues and in the context of the models that students are working with. Here we see great potential for furthering efforts that involve students in evaluating each others’ models and then analyzing any potential harmful biases \cite{morales2024youth}.

Finally, the popularization of large language models and generative models at-large challenges us to think about what to glassbox and blackbox in ways that are appropriate for k-12 students. Williams \cite{williams_popbots_2019}, has experimented with generative models in music, taking a data-driven approach. It is particularly important to develop tools that can help students visualize and explore the training data used by generative models and to visualize how their learning algorithms work so that we can design learning activities that enable youth to explore, modify and create with generative models. 

\section{Conclusion}
Our goal in this conceptual paper is not pit approaches against one another. The approaches, in fact, offer partial but complimentary ways for introducing novices to ML. Glassboxing data, learning algorithms or both provides valuable ways to engage learners in creating ML-powered applications. In conceptualizing these approaches we are reminded that learning and teaching ML must involve attending to functional (the how models work), personal (the how learners can create applications that relate to their interests) and critical (the how learners engage with algorithmic justice) aspects of computing literacies  \cite{kafai2022revaluation}. Regardless of whether we take a data-driven, learning algorithm-driven, or integrative approach we must strive to support these three aspects. We hope that our conceptualization of three emerging approaches contributes to advance our common understanding of ML education in K-12.  

\begin{acks}
This work was supported by NSF grant \#2342438. Views expressed in this paper are those of the authors and do not necessarily reflect the views of NSF or the University of Pennsylvania.

\end{acks}
%%
%% The acknowledgments section is defined using the "acks" environment
%% (and NOT an unnumbered section). This ensures the proper
%% identification of the section in the article metadata, and the
%% consistent spelling of the heading.

%%
%% The next two lines define the bibliography style to be used, and
%% the bibliography file.
\bibliographystyle{ACM-Reference-Format}
\bibliography{sample-base}

\end{document}